\documentclass[%
 reprint,
 amsmath,amssymb,
 aps,
 prapplied,
]{revtex4-2}

\usepackage{gensymb}
\usepackage{braket}
\usepackage{graphicx}
\usepackage{dcolumn}
\usepackage{commath}
\usepackage{bm}

\begin{document}

\preprint{APS/123-QED}

\title{Quantum Coherence is Preserved in Extremely Dispersive Plasmonic Media}

\author{Yury S. Tokpanov}
\author{James S. Fakonas}
\author{Benjamin Vest}
\author{Harry A. Atwater}%
 \email{haa@institution.edu}
\affiliation{%
Thomas J. Watson Laboratories of Applied Physics, California Institute of Technology, Pasadena, California 91125
}%

\date{\today}

\begin{abstract}
Quantum plasmonics experiments have on multiple occasions reported the observation of quantum coherence of discrete plasmons, which exhibit remarkable preservation of quantum interference visibility, a seemingly surprising feature for systems mixing light and matter with high Ohmic losses during propagation. However, most experiments to date used essentially weakly-confined plasmons, which experience limited light-matter hybridization, thus limiting the potential for decoherence. Here, we report quantum coherence of plasmons near the surface plasmon polariton (SPP) resonance frequency, where plasmonic dispersion and confinement is much stronger than in previous experiments. We generated polarization-entangled pairs of photons using spontaneous parametric down conversion and transmitted one of the photons through a plasmonic hole array designed to convert incident single photons into highly-dispersive single SPPs. We find the quality of photon entanglement after the plasmonic channel to be unperturbed by the introduction of a highly dispersive plasmonic element. Our findings provide a lower bound of 100 femtoseconds for the pure dephasing time for dispersive plasmons in gold, and show that even in a highly dispersive regime surface plasmons preserve quantum mechanical correlations, making possible harnessing the power of extreme light confinement for integrated quantum photonics.
\end{abstract}

\maketitle


\section{Introduction}

Understanding the quantum nature of light and matter is of central importance for advancing modern technology.  For example, one approach to physical realization of a quantum computer is envisioned to be through the use of linear optical components \cite{1linoptquant}, which can be arranged in the form of integrated photonic circuits. Since some of the building block elements for such a scalable quantum photonic system (phase shifters, modulators, directional couplers, etc) and the coherence and fabrication requirements are very similar to present-day chip-based nanophotonic circuit elements, recent experimental advances \cite{2nanophasedarray} has inspired optimism about the technical feasibility of quantum integrated photonic systems, if suitable single-photon sources, memories and detectors can be realized.

One of the important branches of photonics is plasmonics, which enables extreme light confinement utilizing surface plasmon polaritons (SPPs), the quanta of so-called surface plasma waves that are excited on the boundary between a metal and a dielectric \cite{3plasmalosses}. SPPs are bosons, so their quantum statistical behavior is expected to be similar to that of photons. By confining electromagnetic energy in small modal volumes, plasmonics allows significantly enhanced light-matter interactions at the nanoscale, and has found interesting applications in classical photonics for sensing \cite{4sers,5spsensors}, sub-diffraction limit imaging \cite{6nearfield,7singleqd}, and paving the way towards strong light-matter interactions, by reaching for example strong coupling regimes \cite{8spporgmol,9strong}. However, light-matter hybridization in SPPs has an important consequence. SPPs are collective excitation of electrons with a mixed electronic and electromagnetic character, while photons in free space are purely electromagnetic excitations, and light propagating through dielectric linear media is described by polaritons, mixing electromagnetic excitation and the motion of bound electrons that do not experience much interactions with the rest of the environment. Through the motion of free electrons in the metal SPPs are coupled to matter by many degrees of freedom. One consequence is that SPPs experience propagating losses due to the Ohmic losses of the moving electrons comprised in an SPP. In fully quantum optical picture, light-matter hybridization could be anticipated to make SPPs sensitive to decoherence from dephasing which is often present in systems with Ohmic losses, leading to disappearance of their quantum features such as entanglement.

Recently, several groups have performed plasmonic analogues to landmark quantum optics experiments using SPPs in lieu of single photons, yielding results such as single plasmon interferences \cite{10spinter, 11duality}, plasmonic Hong-Ou-Mandel experiments \cite{12twoplasmon,13quantinter, 14lossy} and entanglement experiments \cite{15holes,16remote,17path}. These successful experiments faced significant plasmonic absorption, manifested as Ohmic losses, but managed to preserve enough of the plasmons to highlight various quantum features of discrete SPPs. However, an apparently surprising result was that most of performed experiments reported very good or excellent preservation of the quantum interference contrast, possibly indicating that pure dephasing processes are much slower than pure absorption. Notably in each of these experiments, despite the fact that the experimental conditions pre-selected and tested the coherence of the ``surviving" non-absorbed plasmons, the strong plasmonic absorption observed is the undeniable sign of non-negligible coupling between the particles and their environment. The observation of such coupling is an indication that the degrees of freedom of the plasmons are likely to become entangled with the degrees of freedom of the environment, a description that is commonly used to explain the vanishing of quantum interference features, or in other words, decoherence. This has been verified, for example, in an experiment based on plasmonic waveguides \cite{18decoherence}. Decoherence is one of the limiting factors for current and future quantum technology. Hence the question of how SPPs lose quantum mechanical coherence and if their quantum properties can be protected over long propagation distances or under strong light-matter interactions is of significant importance.

We note that until now, quantum plasmonics experiments used exclusively plasmons in a regime far from the SPP resonance. In other words, the plasmons exhibited a highly ``photon-like" behavior with weak confinement, that intrinsically limits the decoherence processes. Indeed, in the case of photon-like plasmons, we expect the plasmon resonances to have only a weak admixture with the electronic degrees of freedom in the metal, leaving the gateway only ajar to significant coupling and entanglement between the SPPs and the metallic environment. Therefore, in this photon-like regime, one could argue that from the perspective of decoherence processes, some quantum plasmonics experiments are somewhat analogous to other quantum optics experiments performed with photons all the way and exhibiting no decoherence. That would also mean that plasmonic losses here play a role that is not different from optical losses introduced by beam splitters, stray reflections, or neutral density filters. Thus, there is need to investigate quantum plasmonics in other regimes of plasmon propagation, where the competition between absorption and pure dephasing could result in observable decoherence.

In this paper, we report results of a quantum plasmonics experiment to investigate the robustness of coherence in a highly dispersive plasmon regime and the disappearance of quantum entanglement. More precisely, in a series of experiments inspired by \cite{15holes}, we measure the preservation of polarization entanglement between two photons after one photon is converted into a plasmon propagating on a hole array, which is then subsequently reconverted into a photon \cite{19eot}. We performed polarization entanglement experiments in plasmonic hole arrays with circular holes which are designed to be in a highly dispersive regime, i.e. with single plasmons close to the SPP resonance. In this highly dispersive regime, SPPs are tightly confined and have a much stronger interaction with the electronic system (one manifestation of which is larger absorption), which in principle can lead to the destruction of quantum correlations. This experiment aims to build a better understanding of the robustness of quantum phenomena in quantum plasmonics.

\section{Results and discussion}
In our experimental work, we generate pairs of polarization-entangled photons, propagating along two different paths, and interpose a plasmonic hole array in the path of one of the photons. This photon is thus converted into a plasmon, and the detected signal consists of plasmons reconverted into free-space photons after plasmon propagation over a few hundred nanometers in the hole array \cite{15holes}. The quality of polarization entanglement between both outcoupled photons is measured and is representative of the effects of light-matter interactions during plasmon propagation. Whereas the work reported in \cite{15holes} probed hole arrays with linear dispersion (close to light line), we use a plasmonic hole array that operates in a highly dispersive regime, to probe plasmon decoherence. We measure preservation of entanglement between photons even in this regime.

\subsection{Generation of entangled pairs of particles}
\begin{figure*}
\includegraphics[width=\textwidth]{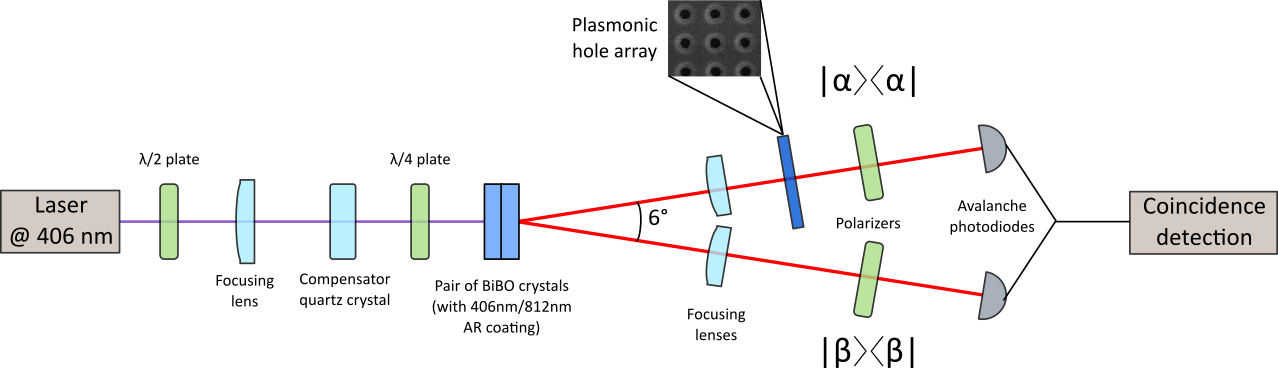}
\caption{\label{fig:setup} Experimental setup for the measurement of polarization-entanglement preservation. Pump photons at 406 nm are sent toward a pair of BiBO crystals and generate pairs of polarization entangled photons that propagate along two separate paths. Along the upper path, we can insert a metallic hole array, and measure the transmission of the entangled light that has been coupled to plasmons.}
\end{figure*}

As a source of polarization-entangled photons we used type-I spontaneous parametric down-conversion (SPDC), occurring in a pair of nonlinear bismuth borate (BiBO) crystals. They are rotated by 90\degree with respect to each other and glued together \cite{20spdc} (Fig.~\ref{fig:setup}), so that one crystal has its axis in the horizontal plane and the other one in a vertical plane. The pair of crystals is pumped by a laser diode emitting at 406 nm. The pump photons are linearly polarized at 45\degree with respect to the nonlinear crystal axis planes, so that type-I SPDC generate pairs of photons at 812 nm that are polarized parallel to either the horizontal direction or the vertical direction with equal probabilities. This setup generates polarization-entangled photons that, before their interaction with the environment $\ket{E}$ (plasmonic sample), can be described by the superposition state:
\begin{eqnarray} \label{eq:psiinit}
\ket{\psi}_{initial}=\frac{1}{\sqrt{2}}[\ket{H,H}+e^{i\Delta\phi_{c}}\ket{V,V}]\otimes\ket{E}],
\end{eqnarray}
where $\Delta\phi_{c}$ is a phase delay between the two polarizations, due to the birefringence of BiBO crystals.

The twin photons propagate in a horizontal plane, along the opposite edges of a cone whose apex angle is 6\degree. Each photon is focused towards a polarizer and a single-photon avalanche diode (SPAD). The detection of a photon by one of the SPADs is a projective measurement of its polarization state. A plasmonic hole array can be placed along one of the propagation paths, thus forcing one of the photon to be temporarily converted into a plasmon before eventually being detected.

In our experiment, in order to correctly estimate the influence of pure dephasing processes, we retain only coincident counts between the two SPADs, i.e. we consider only the case when both photons register counts at the detectors. In other words, when the hole array is in place, we do not record events in which a plasmon has decayed through inelastic interactions with the electronic system -- this is a well-understood mechanism for decoherence. On the contrary, we collect only photons from events in which the plasmon has survived. Such events, in principle, can be affected by elastic interactions or the inner structure of the plasmon quasiparticle.

In the general case, after propagation of the quantum entangled state and before applying any projective measurement, we can consider that the light has become entangled with the environment, and we can write:

\begin{equation} \label{eq:psifin}
\begin{split}
\ket{\psi}_{final}=\frac{1}{\sqrt{|h|^2+|v|^2}} [ &h\ket{H,H}\otimes\ket{E_{H}}+ \\
+&ve^{i\Delta\phi_{c}}\ket{V,V}\otimes\ket{E_{V}} ],
\end{split}
\end{equation} 
where $h$ and $v$ are complex amplitude transmission coefficients for horizontal polarization $\ket{H}$ and vertical polarizations $\ket{V}$ respectively; $\ket{E_H}$ and $\ket{E_V}$ are environmental states, entangled with horizontal and vertical polarizations respectively.

By tracing over environmental states one can obtain a reduced density matrix, from which a probability of a coincidence count can be computed:

\begin{widetext}
\begin{equation} \label{eq:probcc}
\begin{split}
P_{cc}(\alpha,\beta)&=\braket{\alpha,\beta|\hat{\rho}_{reduced}|\alpha,\beta}= \\
&=\frac{1}{1+\frac{|v|^2}{|h|^2}} \left[ \sin^2\alpha \sin^2\beta+\frac{|v|^2}{|h|^2}\cos^2{\alpha} \cos^2{\beta}
+\frac{1}{2}\sin(2\alpha)\sin(2\beta)\frac{|v|}{|h|}\abs{\braket{E_V|E_H}}\cos(\Delta\phi_E+\Delta\phi+\Delta\phi_c) \right],
\end{split}
\end{equation}
\end{widetext}
where $\alpha$ and $\beta$ are polarizers angles (see Fig.1), $\frac{v}{h}=\frac{|v|}{|h|}e^{i\Delta\phi}$, $\Delta\phi$ being the phase difference between the complex amplitudes $v$ and $h$, $\braket{E_V|E_H}=\abs{\braket{E_V|E_H}}e^{i\Delta\phi_E}$, $\phi_E$ being the phase difference between the two environmental states, and $\alpha$ and $\beta$ are the polarizers directions with respect to the vertical axis.

The first two terms can be obtained by classical analysis, whereas the last term is the so-called quantum interference term, which represents quantum mechanical nature of our system. Indeed, Eq. (\ref{eq:psifin}) describes a superposition state. The quantum interference term can be understood as the interference amplitude between the two terms of the superposition state when projective measurements are carried out on the two-particle state. The amplitude of this term depends on several factors. It depends sinusoidally on the polarizers directions, and is maximum for appropriate choices of the polarizer directions verifying $\abs{\sin(2\alpha)\sin(2\beta)}=1$. This corresponds to the situation where the photonic parts of both terms in Eq. (\ref{eq:psifin}) are projected on a common state with equal amplitude. The amplitude of the quantum term is partially governed by the ratio $\frac{v}{h}=\frac{|v|}{|h|}e^{i(\Delta\phi+\Delta\phi_c)}$, which includes all perturbations inherent to the setup that affect the balance between the horizontal and the vertical polarization. Finally, we note here that the magnitude of quantum interference is also determined by the overlap between different environment states $\braket{E_V|E_H}=\abs{\braket{E_V|E_H}}e^{i\Delta\phi_E}$, which represents quantum mechanical decoherence. The presence of $\Delta\phi_c$ in the last cosine factor of the quantum interference term shows that, in order to make judgements about quantum decoherence, one has to take a great care in eliminating or measuring phase differences between different polarizations. This can be done by inserting another birefringent element in the setup that will compensate the phase difference between the two polarizations. Optimization and alignment of our SPDC source included tweaking of a $\lambda/4$ plate (Fig. 1), which allowed us to experimentally eliminate $\Delta\phi_c$ in Eq. (\ref{eq:probcc}).

We now consider $h=v$, which represents equal probability of detecting horizontally or vertically polarized pairs of photons and is fulfilled when using circular hole array. In the case of perfect coherence $\braket{E_V|E_H}=1$ $(E_V=E_H)$ we get the rather simple expression $P_{cc}(\alpha,\beta)=\frac{1}{2}\cos^2(\alpha-\beta)$. There is no entanglement between the photon state and the environment. This ensures the preservation of polarization entanglement between photons. By contrast, in the case of total decoherence $\braket{E_V|E_H}=0$ we get a constant probability $P_{cc}(\alpha,\beta=45\degree)=\frac{1}{4}$ regardless of $\alpha$ (if $\beta$ is kept fixed at 45\degree). Both terms of the superposition in (\ref{eq:psifin}) are now incoherent, and quantum interferences vanish. The measured state can be considered as a statistical mixture of the two states $\ket{H,H}$ and $\ket{V,V}$ in equal proportions.

These considerations suggest a measure of quality of the entanglement, visibility, which we define simply as the visibility of the cosine curve described by $P_{cc}(\alpha,\beta=45\degree)$ for the case when we keep polarizer $\beta$ fixed at 45\degree (both polarizations contribute to the measurement) $V=\frac{P^{max}_{cc}-P^{min}_{cc}}{P^{max}_{cc}+P^{min}_{cc}}$, where $P^{min}_{cc}$ is the minimum probability of coincidence count (rate in an experiment) and $P^{max}_{cc}$ is the maximum count rate. For $h=v$ and $\Delta\phi_c=0\degree$ visibility is equal to $V=\abs{\braket{E_V|E_H}}\cos(\Delta\phi_E)$. From the above analysis we get that $V=100\%$ for fully entangled (quantum) light ($\braket{E_V|E_H}=1$), and $V=0\%$ for for a pure statistical mixture of polarizations (``classical'' light, $\braket{E_V|E_H}=0$). Note, that visibility of a cosine $P_{cc}(\alpha,\beta=45\degree)$ is identical to the visibility of the cosine $P_{cc}(\alpha,\beta=135\degree)$, hence we can use either one of them, or use one versus another to validate the correctness of the measurement.

In addition to that, we performed Bell's inequalities violation measurements, where we use Bell's inequalities in so-called CHSH form \cite{21hiddenvariables,22chsh}. We performed 16-point measurements in order to calculate Bell's parameter $S$, comparing our experimental measurement with the best possible prediction of any classical local hidden variable theory (LHVT). $S>2$ indicates the impossibility of the explanation by any LHVT.

\begin{figure}
\includegraphics[width=\linewidth]{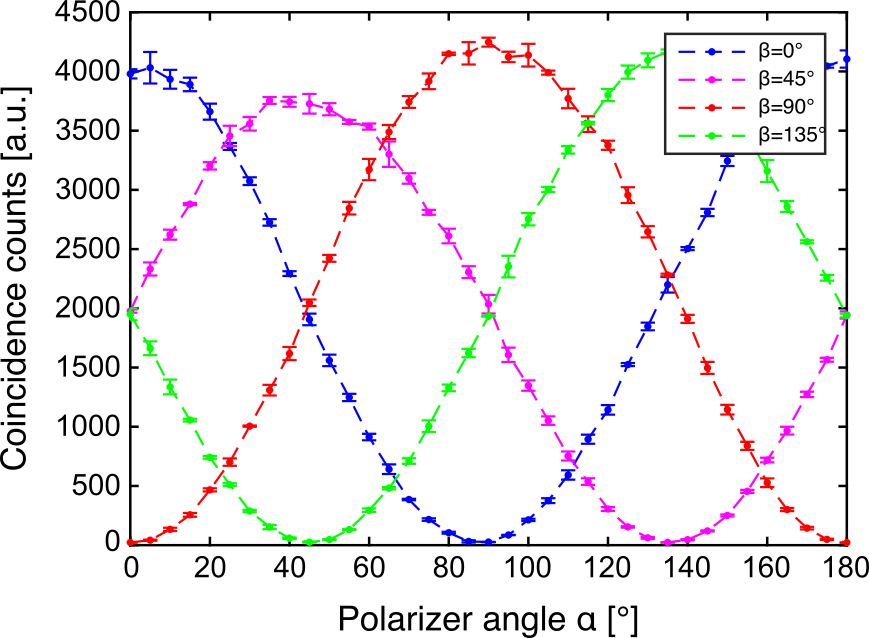}
\caption{\label{fig:calibration}Calibration of the setup, and entanglement between produced pairs of photons. Number of coincidence counts as a function of polarizers angles without plasmonic sample (solid lines are fits to cosine). The visibility of the different cosine fits is nearly equal to one, indicating quasi-perfect entanglement between the photons of our SPDC source.}
\end{figure}

We characterized our SPDC source (Fig.~\ref{fig:calibration}) without plasmonic samples, measuring visibility on the order of $V=99\%\pm1\%$ and $S=2.81\pm0.02$, which is just a standard deviation away from the maximal theoretical value $S_{max}=2\sqrt{2}\approx 2.83$. From this, we conclude that we have high quality pairs of entangled photons. In the next section, we investigate the influence of the insertion of a plasmonic hole array on the quality of entanglement between photons, as defined by the previous measurement procedures.

\subsection{Hole array with nonlinear dispersion}

In this section, we investigate the same polarization entanglement process using this time single SPPs that propagate in a hole array, and most importantly in a regime of highly nonlinear dispersion, far from the light line, in an attempt to reveal effects of pure dephasing on decoherence through a decrease of entanglement visibility.  In the highly dispersive regime, the quasiparticle confinement at the metal/dielectric interface is much stronger due to the reduction in the plasmon wavelength. In other words, the plasmons excitation wavefunction has a much larger overlap with the electronic degrees of freedom in the metal. As a consequence, plasmons in the highly dispersive regime have a generally higher rate of interaction with the electronic system than in the case of materials with linear dispersion. Hence, in addition to shorter total decoherence time, one can expect a shorter pure dephasing time $T^{*}_2$ (which is the relevant time scale probed by our experiment). The group velocity of highly dispersive plasmons can be an order of magnitude smaller than for plasmons in the photon-like regime, so that these plasmons propagate for a longer time $t_p$ (even if the propagation distances are the same). Therefore, for comparable experiments (i.e. similar propagation distances), strongly-confined plasmons are expected to experience greater decoherence thus exhibit weaker quantum interference than photon-like plasmons.

As a warm-up experiment and to be able to compare the influence of the non-linearity of the dispersion regime with the linear regime, we performed a series of experiments with hole arrays and plasmons propagating at the gold-air interface, the structure being optimized to operate in the linear (``photon-like") dispersion regime. Results are shown in Supplemental Materials and we found no evidence of decoherence when close to the light line, as was frequently reported until now \cite{supplemental1}.

\begin{figure}
\includegraphics[width=\linewidth]{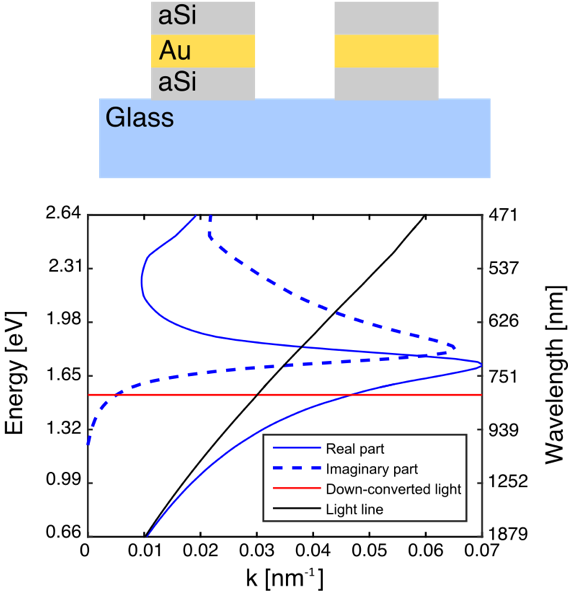}
\caption{\label{fig:dispersion}Plasmonic hole array design in our experiment : cross-sectional schematic (top) and dispersion relation of circular hole arrays for SPPs supported in the silicon/gold/silicon structure, which exhibits strongly nonlinear dispersion at 812 nm (bottom). The red line shows the energy of the plasmons excited in our experiment.}
\end{figure}

\begin{figure}
\includegraphics[width=\linewidth]{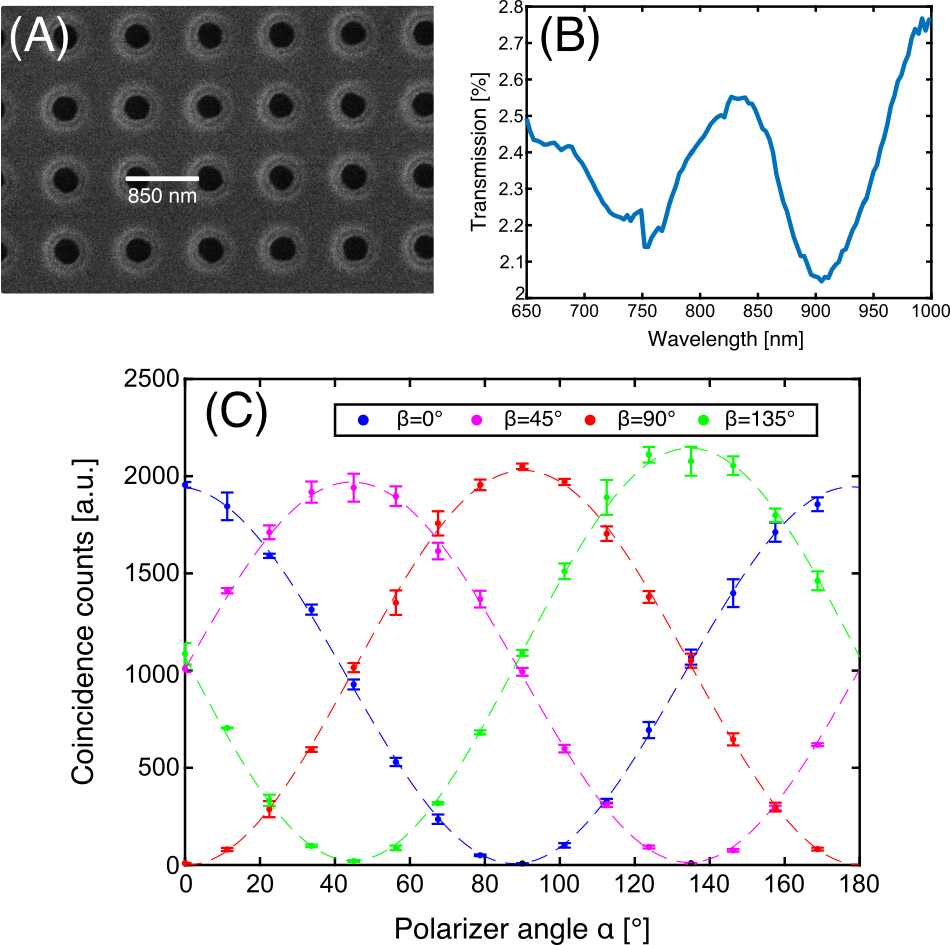}
\caption{\label{fig:result}Hole array for plasmons in higly-dispersive regime. (A) SEM image of the sample. The period of the 2D array is 850 nm. The different material species experience different milling rate that slightly affect the shape of the holes perimeter. (B) Transmission spectrum of the hole array. The holes being circular, there is no polarization dependence. The broadening of the transmission feature around 810 nm can be attributed to the imperfect shape of the holes. Note that the plasmons experience a significant absorption. (C) Number of coincidence counts as a function of polarizer angles in the presence of the hole array with highly-dispersive regime (solid lines are fits to cosine). Whatever the choice of $\beta$ is, and even when placed at 45\degree or 135\degree, the visibility of quantum interference remains almost equal to one, indicating near perfect preservation of entanglement between particles.}
\end{figure}

In order to probe the highly dispersive regime in an analogous situation, we excite plasmons at the interface between gold and amorphous silicon. Amorphous silicon has a higher dielectric constant than glass, moving the SPP resonance frequency close to the frequency of entangled photons (see Fig.~\ref{fig:dispersion}). The use of a higher index material leads both to a stronger confinement of the plasmons, with a 6-fold increase of the plasmon wave vector and a 12-fold reduction of their group velocity compared with plasmons propagating at the interface between gold and air ($0.05c$ versus $0.59c$, where $c$ is the speed of light in vacuum). An initial choice for the range of geometrical parameters (hole dimensions, array periodicity, films thicknesses) was made after performing finite-difference time-domain simulations in Lumerical software of the structure designed to enhance extraordinary transmission at 812 nm -- the wavelength of our down-converted photons. Subsequently, we fabricated a number of hole arrays with different geometrical parameters within this range to experimentally find the structure, that achieves target performance. This procedure allowed us to identify and fabricate a sample with optimal design, which is a 2 mm $\times$ 2 mm hole array configured in a three-layer structure (50 nm of amorphous silicon -- 100 nm of gold -- 50 nm of amorphous silicon) and a periodicity $P=850\textrm{nm}$ (Fig.~\ref{fig:result}(A)). We find a plasmon-enhanced transmission peak at the desired 812 nm wavelength (Fig.~\ref{fig:result}(B)). In this hole array SPPs excited on the top and bottom gold surfaces are uncoupled and have the same dispersion. We performed additional analysis and confirmed that the fabricated structure correctly reproduces the initially simulated behavior and operates in the non-linear dispersion regime, far from the light line \cite{supplemental2}.

Thanks to the rather large size of the sample we were able to collect a large portion of transmitted light and hence improve our statistics, even in presence of considerable losses and with intrinsically low transmission (Fig.~\ref{fig:result}(C)). We recorded a visibility of $V=98\%\pm2\%$ and Bell's number $S=2.83\pm0.04$: this measurement implies that even in the highly-dispersive regime, the entanglement is perfectly preserved and no quantitative signs of pure dephasing could be detected. Through numerical computation of the dispersion relation, we can estimate the propagation time to be on the order of $ t_p \sim \frac{1.2 \mu \textrm{m}}{0.05c} \sim 80 \, \textrm{fs}$, based on the distance between two diagonally separated holes (eigenmodes of a hole array are diagonally oriented). This time is much longer than the value of total dephasing time $T_2=2T_1=20 \, \textrm{fs}$, that can be estimated from the absorption length in our sample (less than 200 nm, in agreement with literature reported values \cite{23dephasing}). Another approach towards an estimation of the propagation time, based on an approximate lorentzian fit of the resonance experienced by the plasmons, gives a similar time. We therefore conclude, that in our system pure dephasing is a remarkably slow process compared to absorption, and the order of magnitude of the lower bound of the dephasing time can be estimated to be 100 fs, which is similar to the value reported in \cite{18decoherence}. 

We note, however, that this time could be in practice much higher, as our experiment remarkably reports no quantitative trace of quantum decoherence. Performing the same experiment in an even more highly dispersive regime could hypothetically allow us to make decoherence process eventually visible at some extent. This would make the degradation of the fringes visibility capable of being modeled, one of the parameters of such a model being the pure dephasing time, that could be more precisely estimated. The design of such an experiment is however fundamentally limited by the high level of absorption that coexists with the enhancement of light-matter interactions. It dramatically reduces the signal level at the output of the plasmonic path progressively to almost zero.

In summary, we have examined the influence of plasmon dispersion on the quantum decoherence properties of surface plasmons. The excitation of highly-dispersive plasmons did not result in the reduction of the quality of a single-particle quantum state for transmitted light. Plasmons excited in hole arrays are found to preserve quantum mechanical correlations, even in the presence of extreme dispersion near the plasmon resonance and strong absorption. Moreover, the focus of our measurements is on the elastic dephasing processes, which consideration is commonly neglected in quantum optics modeling due to its supposed insignificance. Our findings provide experimental proof for such an assumption, and also emphasize the difference between decay and decoherence.  Thus, we conclude that despite being lossy, plasmonic structures may find applications in the realms of quantum technology, where the power of extreme light confinement can be effectively leveraged.

\section*{Author contributions}
Y.S.T., J.S.F. and H.A.A. proposed the original idea. Y.S.T. realized all experiments and data analysis. B. V. helped in discussion. Y.S.T., B.V., H.A.A. wrote the paper, and all authors discussed and revised the manuscript.

\section*{Acknowledgments}
We acknowledge Yousif Kelaita, Artur Davoyan, Ruzan Sokhoyan, Ragip Pala, Dagny Fleischman, Zachary Aitken, Sunita Darbe for help with equipment training and scientific advice.

This work was supported by the Air Force Office of Scientific Research under grant number FA9550-16-1-0019.
The authors declare no competing interests.



\begin{thebibliography}{25}%
\makeatletter
\providecommand \@ifxundefined [1]{%
 \@ifx{#1\undefined}
}%
\providecommand \@ifnum [1]{%
 \ifnum #1\expandafter \@firstoftwo
 \else \expandafter \@secondoftwo
 \fi
}%
\providecommand \@ifx [1]{%
 \ifx #1\expandafter \@firstoftwo
 \else \expandafter \@secondoftwo
 \fi
}%
\providecommand \natexlab [1]{#1}%
\providecommand \enquote  [1]{``#1''}%
\providecommand \bibnamefont  [1]{#1}%
\providecommand \bibfnamefont [1]{#1}%
\providecommand \citenamefont [1]{#1}%
\providecommand \href@noop [0]{\@secondoftwo}%
\providecommand \href [0]{\begingroup \@sanitize@url \@href}%
\providecommand \@href[1]{\@@startlink{#1}\@@href}%
\providecommand \@@href[1]{\endgroup#1\@@endlink}%
\providecommand \@sanitize@url [0]{\catcode `\\12\catcode `\$12\catcode
  `\&12\catcode `\#12\catcode `\^12\catcode `\_12\catcode `\%12\relax}%
\providecommand \@@startlink[1]{}%
\providecommand \@@endlink[0]{}%
\providecommand \url  [0]{\begingroup\@sanitize@url \@url }%
\providecommand \@url [1]{\endgroup\@href {#1}{\urlprefix }}%
\providecommand \urlprefix  [0]{URL }%
\providecommand \Eprint [0]{\href }%
\providecommand \doibase [0]{https://doi.org/}%
\providecommand \selectlanguage [0]{\@gobble}%
\providecommand \bibinfo  [0]{\@secondoftwo}%
\providecommand \bibfield  [0]{\@secondoftwo}%
\providecommand \translation [1]{[#1]}%
\providecommand \BibitemOpen [0]{}%
\providecommand \bibitemStop [0]{}%
\providecommand \bibitemNoStop [0]{.\EOS\space}%
\providecommand \EOS [0]{\spacefactor3000\relax}%
\providecommand \BibitemShut  [1]{\csname bibitem#1\endcsname}%
\let\auto@bib@innerbib\@empty
\bibitem [{\citenamefont {Knill}\ \emph {et~al.}(2001)\citenamefont {Knill},
  \citenamefont {Laflamme},\ and\ \citenamefont {Milburn}}]{1linoptquant}%
  \BibitemOpen
  \bibfield  {author} {\bibinfo {author} {\bibfnamefont {E.}~\bibnamefont
  {Knill}}, \bibinfo {author} {\bibfnamefont {R.}~\bibnamefont {Laflamme}},\
  and\ \bibinfo {author} {\bibfnamefont {G.~J.}\ \bibnamefont {Milburn}},\
  }\bibfield  {title} {\bibinfo {title} {A scheme for efficient quantum
  computation with linear optics},\ }\href@noop {} {\bibfield  {journal}
  {\bibinfo  {journal} {Nature}\ }\textbf {\bibinfo {volume} {409}},\ \bibinfo
  {pages} {46} (\bibinfo {year} {2001})}\BibitemShut {NoStop}%
\bibitem [{\citenamefont {Sun}\ \emph {et~al.}(2013)\citenamefont {Sun},
  \citenamefont {Timurdogan}, \citenamefont {Yaacobi}, \citenamefont
  {Hosseini},\ and\ \citenamefont {Watts}}]{2nanophasedarray}%
  \BibitemOpen
  \bibfield  {author} {\bibinfo {author} {\bibfnamefont {J.}~\bibnamefont
  {Sun}}, \bibinfo {author} {\bibfnamefont {E.}~\bibnamefont {Timurdogan}},
  \bibinfo {author} {\bibfnamefont {A.}~\bibnamefont {Yaacobi}}, \bibinfo
  {author} {\bibfnamefont {E.~S.}\ \bibnamefont {Hosseini}},\ and\ \bibinfo
  {author} {\bibfnamefont {M.~R.}\ \bibnamefont {Watts}},\ }\bibfield  {title}
  {\bibinfo {title} {Large-scale nanophotonic phased array},\ }\href@noop {}
  {\bibfield  {journal} {\bibinfo  {journal} {Nature}\ }\textbf {\bibinfo
  {volume} {493}},\ \bibinfo {pages} {195} (\bibinfo {year}
  {2013})}\BibitemShut {NoStop}%
\bibitem [{\citenamefont {Ritchie}(1957)}]{3plasmalosses}%
  \BibitemOpen
  \bibfield  {author} {\bibinfo {author} {\bibfnamefont {R.~H.}\ \bibnamefont
  {Ritchie}},\ }\bibfield  {title} {\bibinfo {title} {Plasma losses by fast
  electrons in thin films},\ }\href@noop {} {\bibfield  {journal} {\bibinfo
  {journal} {Physical Review}\ }\textbf {\bibinfo {volume} {106}},\ \bibinfo
  {pages} {874} (\bibinfo {year} {1957})}\BibitemShut {NoStop}%
\bibitem [{\citenamefont {Alvarez-Puebla}\ \emph {et~al.}(2007)\citenamefont
  {Alvarez-Puebla}, \citenamefont {B.~Cui}, \citenamefont {Veres},\ and\
  \citenamefont {Fenniri}}]{4sers}%
  \BibitemOpen
  \bibfield  {author} {\bibinfo {author} {\bibfnamefont {R.}~\bibnamefont
  {Alvarez-Puebla}}, \bibinfo {author} {\bibfnamefont {B.}~\bibnamefont
  {Cui}}, \bibinfo {author} {\bibfnamefont {J.-~P.}\
  \bibnamefont {Bravo-Vasquez}}, \bibinfo {author} {\bibfnamefont {T.}~\bibnamefont
  {Veres}},\ and\ \bibinfo {author} {\bibfnamefont {H.}~\bibnamefont
  {Fenniri}},\ }\bibfield  {title} {\bibinfo {title} {Nanoimprinted SERS-active
  substrates with tunable surface plasmon resonances},\ }\href@noop {}
  {\bibfield  {journal} {\bibinfo  {journal} {The Journal of Physical Chemistry
  C}\ }\textbf {\bibinfo {volume} {111}},\ \bibinfo {pages} {6720} (\bibinfo
  {year} {2007})}\BibitemShut {NoStop}%
\bibitem [{\citenamefont {Homola}(2008)}]{5spsensors}%
  \BibitemOpen
  \bibfield  {author} {\bibinfo {author} {\bibfnamefont {J.}~\bibnamefont
  {Homola}},\ }\bibfield  {title} {\bibinfo {title} {Surface plasmon resonance
  sensors for detection of chemical and biological species},\ }\href@noop {}
  {\bibfield  {journal} {\bibinfo  {journal} {Chemical reviews}\ }\textbf
  {\bibinfo {volume} {108}},\ \bibinfo {pages} {462} (\bibinfo {year}
  {2008})}\BibitemShut {NoStop}%
\bibitem [{\citenamefont {Taubner}\ \emph {et~al.}(2006)\citenamefont
  {Taubner}, \citenamefont {Korobkin}, \citenamefont {Urzhumov}, \citenamefont
  {Shvets},\ and\ \citenamefont {Hillenbrand}}]{6nearfield}%
  \BibitemOpen
  \bibfield  {author} {\bibinfo {author} {\bibfnamefont {T.}~\bibnamefont
  {Taubner}}, \bibinfo {author} {\bibfnamefont {D.}~\bibnamefont {Korobkin}},
  \bibinfo {author} {\bibfnamefont {Y.}~\bibnamefont {Urzhumov}}, \bibinfo
  {author} {\bibfnamefont {G.}~\bibnamefont {Shvets}},\ and\ \bibinfo {author}
  {\bibfnamefont {R.}~\bibnamefont {Hillenbrand}},\ }\bibfield  {title}
  {\bibinfo {title} {Near-field microscopy through a sic superlens},\
  }\href@noop {} {\bibfield  {journal} {\bibinfo  {journal} {Science}\ }\textbf
  {\bibinfo {volume} {313}},\ \bibinfo {pages} {1595} (\bibinfo {year}
  {2006})}\BibitemShut {NoStop}%
\bibitem [{\citenamefont {Farahani}\ \emph {et~al.}(2005)\citenamefont
  {Farahani}, \citenamefont {Pohl}, \citenamefont {Eisler},\ and\ \citenamefont
  {Hecht.}}]{7singleqd}%
  \BibitemOpen
  \bibfield  {author} {\bibinfo {author} {\bibfnamefont {J.~N.}\ \bibnamefont
  {Farahani}}, \bibinfo {author} {\bibfnamefont {D.~W.}\ \bibnamefont {Pohl}},
  \bibinfo {author} {\bibfnamefont {H.-J.}\ \bibnamefont {Eisler}},\ and\
  \bibinfo {author} {\bibfnamefont {B.}~\bibnamefont {Hecht.}},\ }\bibfield
  {title} {\bibinfo {title} {Single quantum dot coupled to a scanning optical
  antenna: a tunable superemitter},\ }\href@noop {} {\bibfield  {journal}
  {\bibinfo  {journal} {Phys. Rev. Lett.}\ }\textbf {\bibinfo {volume} {95}},\
  \bibinfo {pages} {17402} (\bibinfo {year} {2005})}\BibitemShut {NoStop}%
\bibitem [{\citenamefont {Dintinger}\ \emph {et~al.}(2005)\citenamefont
  {Dintinger}, \citenamefont {Klein}, \citenamefont {Bustos}, \citenamefont
  {Barnes},\ and\ \citenamefont {Ebbesen}}]{8spporgmol}%
  \BibitemOpen
  \bibfield  {author} {\bibinfo {author} {\bibfnamefont {J.}~\bibnamefont
  {Dintinger}}, \bibinfo {author} {\bibfnamefont {S.}~\bibnamefont {Klein}},
  \bibinfo {author} {\bibfnamefont {F.}~\bibnamefont {Bustos}}, \bibinfo
  {author} {\bibfnamefont {W.~L.}\ \bibnamefont {Barnes}},\ and\ \bibinfo
  {author} {\bibfnamefont {T.~W.}\ \bibnamefont {Ebbesen}},\ }\bibfield
  {title} {\bibinfo {title} {Strong coupling between surface plasmon-polaritons
  and organic molecules in subwavelength hole arrays},\ }\href@noop {}
  {\bibfield  {journal} {\bibinfo  {journal} {Physical Review B}\ }\textbf
  {\bibinfo {volume} {71}},\ \bibinfo {pages} {035424} (\bibinfo {year}
  {2005})}\BibitemShut {NoStop}%
\bibitem [{\citenamefont {Chang}\ \emph {et~al.}(2007)\citenamefont {Chang},
  \citenamefont {Sorensen}, \citenamefont {Hemmer},\ and\ \citenamefont
  {Lukin}}]{9strong}%
  \BibitemOpen
  \bibfield  {author} {\bibinfo {author} {\bibfnamefont {D.~E.}\ \bibnamefont
  {Chang}}, \bibinfo {author} {\bibfnamefont {A.~S.}\ \bibnamefont
  {Sorensen}}, \bibinfo {author} {\bibfnamefont {P.~R.}\ \bibnamefont
  {Hemmer}},\ and\ \bibinfo {author} {\bibfnamefont {M.~D.}\ \bibnamefont
  {Lukin}},\ }\bibfield  {title} {\bibinfo {title} {Strong coupling of single
  emitters to surface plasmons},\ }\href@noop {} {\bibfield  {journal}
  {\bibinfo  {journal} {Physical Review B}\ }\textbf {\bibinfo {volume} {76}},\
  \bibinfo {pages} {035420} (\bibinfo {year} {2007})}\BibitemShut {NoStop}%
\bibitem [{\citenamefont {Dheur}\ \emph {et~al.}(2016)\citenamefont {Dheur},
  \citenamefont {Devaux}, \citenamefont {Ebbesen}, \citenamefont {Baron},
  \citenamefont {Rodier}, \citenamefont {Hugonin}, \citenamefont {Greffet},
  \citenamefont {Messin},\ and\ \citenamefont {Marquier}}]{10spinter}%
  \BibitemOpen
  \bibfield  {author} {\bibinfo {author} {\bibfnamefont {M.-C.}\ \bibnamefont
  {Dheur}}, \bibinfo {author} {\bibfnamefont {E.}~\bibnamefont {Devaux}},
  \bibinfo {author} {\bibfnamefont {T.~W.}\ \bibnamefont {Ebbesen}}, \bibinfo
  {author} {\bibfnamefont {A.}~\bibnamefont {Baron}}, \bibinfo {author}
  {\bibfnamefont {J.-C.}\ \bibnamefont {Rodier}}, \bibinfo {author}
  {\bibfnamefont {J.-P.}\ \bibnamefont {Hugonin}}, \bibinfo {author}
  {\bibfnamefont {J.-J.}\ \bibnamefont {Greffet}}, \bibinfo {author}
  {\bibfnamefont {G.}~\bibnamefont {Messin}},\ and\ \bibinfo {author}
  {\bibfnamefont {F.}~\bibnamefont {Marquier}},\ }\bibfield  {title} {\bibinfo
  {title} {Single-plasmon interferences},\ }\href@noop {} {\bibfield  {journal}
  {\bibinfo  {journal} {Science advances}\ }\textbf {\bibinfo {volume} {2}},\
  \bibinfo {pages} {e1501574} (\bibinfo {year} {2016})}\BibitemShut {NoStop}%
\bibitem [{\citenamefont {Kolesov}\ \emph {et~al.}(2009)\citenamefont
  {Kolesov}, \citenamefont {Grotz}, \citenamefont {Balasubramanian},
  \citenamefont {Stöhr}, \citenamefont {Nicolet}, \citenamefont {Hemmer},
  \citenamefont {Jelezko},\ and\ \citenamefont {Wrachtrup}}]{11duality}%
  \BibitemOpen
  \bibfield  {author} {\bibinfo {author} {\bibfnamefont {R.}~\bibnamefont
  {Kolesov}}, \bibinfo {author} {\bibfnamefont {B.}~\bibnamefont {Grotz}},
  \bibinfo {author} {\bibfnamefont {G.}~\bibnamefont {Balasubramanian}},
  \bibinfo {author} {\bibfnamefont {R.~J.}\ \bibnamefont {Stöhr}}, \bibinfo
  {author} {\bibfnamefont {A.~A.}\ \bibnamefont {Nicolet}}, \bibinfo {author}
  {\bibfnamefont {P.~R.}\ \bibnamefont {Hemmer}}, \bibinfo {author}
  {\bibfnamefont {F.}~\bibnamefont {Jelezko}},\ and\ \bibinfo {author}
  {\bibfnamefont {J.}~\bibnamefont {Wrachtrup}},\ }\bibfield  {title} {\bibinfo
  {title} {Wave-particle duality of single surface plasmon polaritons},\
  }\href@noop {} {\bibfield  {journal} {\bibinfo  {journal} {Nature Physics}\
  }\textbf {\bibinfo {volume} {5}},\ \bibinfo {pages} {470} (\bibinfo {year}
  {2009})}\BibitemShut {NoStop}%
\bibitem [{\citenamefont {Fakonas}\ \emph {et~al.}(2014)\citenamefont
  {Fakonas}, \citenamefont {Lee}, \citenamefont {Kelaita},\ and\ \citenamefont
  {Atwater}}]{12twoplasmon}%
  \BibitemOpen
  \bibfield  {author} {\bibinfo {author} {\bibfnamefont {J.~S.}\ \bibnamefont
  {Fakonas}}, \bibinfo {author} {\bibfnamefont {H.}~\bibnamefont {Lee}},
  \bibinfo {author} {\bibfnamefont {Y.~A.}\ \bibnamefont {Kelaita}},\ and\
  \bibinfo {author} {\bibfnamefont {H.~A.}\ \bibnamefont {Atwater}},\
  }\bibfield  {title} {\bibinfo {title} {Two-plasmon quantum interference},\
  }\href@noop {} {\bibfield  {journal} {\bibinfo  {journal} {Nature Photonics}\
  }\textbf {\bibinfo {volume} {8}},\ \bibinfo {pages} {317} (\bibinfo {year}
  {2014})}\BibitemShut {NoStop}%
\bibitem [{\citenamefont {Martino}\ \emph {et~al.}(2014)\citenamefont
  {Martino}, \citenamefont {Sonnefraud}, \citenamefont {Tame}, \citenamefont
  {Kéna-Cohen}, \citenamefont {Dieleman}, \citenamefont {Özdemir},
  \citenamefont {Kim},\ and\ \citenamefont {Maier}}]{13quantinter}%
  \BibitemOpen
  \bibfield  {author} {\bibinfo {author} {\bibfnamefont {G.~D.}\ \bibnamefont
  {Martino}}, \bibinfo {author} {\bibfnamefont {Y.}~\bibnamefont {Sonnefraud}},
  \bibinfo {author} {\bibfnamefont {M.}~\bibnamefont {Tame}}, \bibinfo {author}
  {\bibfnamefont {S.}~\bibnamefont {Kéna-Cohen}}, \bibinfo {author}
  {\bibfnamefont {F.}~\bibnamefont {Dieleman}}, \bibinfo {author}
  {\bibfnamefont {S.~K.}\ \bibnamefont {Özdemir}}, \bibinfo {author}
  {\bibfnamefont {M.~S.}\ \bibnamefont {Kim}},\ and\ \bibinfo {author}
  {\bibfnamefont {S.~A.}\ \bibnamefont {Maier}},\ }\bibfield  {title} {\bibinfo
  {title} {Observation of quantum interference in the plasmonic hong-ou-mandel
  effect},\ }\href@noop {} {\bibfield  {journal} {\bibinfo  {journal} {Physical
  Review Applied}\ }\textbf {\bibinfo {volume} {1}},\ \bibinfo {pages} {034004}
  (\bibinfo {year} {2014})}\BibitemShut {NoStop}%
\bibitem [{\citenamefont {Vest}\ \emph {et~al.}(2017)\citenamefont {Vest},
  \citenamefont {Dheur}, \citenamefont {Devaux}, \citenamefont {Baron},
  \citenamefont {Rousseau}, \citenamefont {Hugonin}, \citenamefont {Greffet},
  \citenamefont {Messin},\ and\ \citenamefont {Marquier}}]{14lossy}%
  \BibitemOpen
  \bibfield  {author} {\bibinfo {author} {\bibfnamefont {B.}~\bibnamefont
  {Vest}}, \bibinfo {author} {\bibfnamefont {M.-C.}\ \bibnamefont {Dheur}},
  \bibinfo {author} {\bibfnamefont {E.}~\bibnamefont {Devaux}}, \bibinfo
  {author} {\bibfnamefont {A.}~\bibnamefont {Baron}}, \bibinfo {author}
  {\bibfnamefont {E.}~\bibnamefont {Rousseau}}, \bibinfo {author}
  {\bibfnamefont {J.-P.}\ \bibnamefont {Hugonin}}, \bibinfo {author}
  {\bibfnamefont {J.-J.}\ \bibnamefont {Greffet}}, \bibinfo {author}
  {\bibfnamefont {G.}~\bibnamefont {Messin}},\ and\ \bibinfo {author}
  {\bibfnamefont {F.}~\bibnamefont {Marquier}},\ }\bibfield  {title} {\bibinfo
  {title} {Anti-coalescence of bosons on a lossy beam splitter},\ }\href@noop
  {} {\bibfield  {journal} {\bibinfo  {journal} {Science}\ }\textbf {\bibinfo
  {volume} {356}},\ \bibinfo {pages} {1373} (\bibinfo {year}
  {2017})}\BibitemShut {NoStop}%
\bibitem [{\citenamefont {Altewischer}\ \emph {et~al.}(2002)\citenamefont
  {Altewischer}, \citenamefont {Exter},\ and\ \citenamefont
  {Woerdman}}]{15holes}%
  \BibitemOpen
  \bibfield  {author} {\bibinfo {author} {\bibfnamefont {E.}~\bibnamefont
  {Altewischer}}, \bibinfo {author} {\bibfnamefont {M.~P.}\ \bibnamefont
  {Exter}},\ and\ \bibinfo {author} {\bibfnamefont {J.~P.}\ \bibnamefont
  {Woerdman}},\ }\bibfield  {title} {\bibinfo {title} {Plasmon-assisted
  transmission of entangled photons},\ }\href@noop {} {\bibfield  {journal}
  {\bibinfo  {journal} {Nature}\ }\textbf {\bibinfo {volume} {418}},\ \bibinfo
  {pages} {304} (\bibinfo {year} {2002})}\BibitemShut {NoStop}%
\bibitem [{\citenamefont {Dheur}\ \emph {et~al.}(2017)\citenamefont {Dheur},
  \citenamefont {Vest}, \citenamefont {Devaux}, \citenamefont {Baron},
  \citenamefont {Hugonin}, \citenamefont {Greffet}, \citenamefont {Messin},\
  and\ \citenamefont {Marquier}}]{16remote}%
  \BibitemOpen
  \bibfield  {author} {\bibinfo {author} {\bibfnamefont {M.-C.}\ \bibnamefont
  {Dheur}}, \bibinfo {author} {\bibfnamefont {B.}~\bibnamefont {Vest}},
  \bibinfo {author} {\bibfnamefont {E.}~\bibnamefont {Devaux}}, \bibinfo
  {author} {\bibfnamefont {A.}~\bibnamefont {Baron}}, \bibinfo {author}
  {\bibfnamefont {J.-P.}\ \bibnamefont {Hugonin}}, \bibinfo {author}
  {\bibfnamefont {J.-J.}\ \bibnamefont {Greffet}}, \bibinfo {author}
  {\bibfnamefont {G.}~\bibnamefont {Messin}},\ and\ \bibinfo {author}
  {\bibfnamefont {F.}~\bibnamefont {Marquier}},\ }\bibfield  {title} {\bibinfo
  {title} {Remote preparation of single-plasmon states},\ }\href@noop {}
  {\bibfield  {journal} {\bibinfo  {journal} {Physical Review B}\ }\textbf
  {\bibinfo {volume} {96}},\ \bibinfo {pages} {045432} (\bibinfo {year}
  {2017})}\BibitemShut {NoStop}%
\bibitem [{\citenamefont {Fakonas}\ \emph {et~al.}(2015)\citenamefont
  {Fakonas}, \citenamefont {Mitskovets},\ and\ \citenamefont
  {Atwater}}]{17path}%
  \BibitemOpen
  \bibfield  {author} {\bibinfo {author} {\bibfnamefont {J.~S.}\ \bibnamefont
  {Fakonas}}, \bibinfo {author} {\bibfnamefont {A.}~\bibnamefont
  {Mitskovets}},\ and\ \bibinfo {author} {\bibfnamefont {H.~A.}\ \bibnamefont
  {Atwater}},\ }\bibfield  {title} {\bibinfo {title} {Path-entanglement of
  surface plasmons},\ }\href@noop {} {\bibfield  {journal} {\bibinfo  {journal}
  {New Journal of Physics}\ }\textbf {\bibinfo {volume} {17}} (\bibinfo {year}
  {2015})}\BibitemShut {NoStop}%
\bibitem [{\citenamefont {Dlamini}\ \emph {et~al.}(2018)\citenamefont
  {Dlamini}, \citenamefont {Francis}, \citenamefont {Zhang}, \citenamefont
  {Özdemir}, \citenamefont {Chormaic}, \citenamefont {Petruccione},\ and\
  \citenamefont {Tame}}]{18decoherence}%
  \BibitemOpen
  \bibfield  {author} {\bibinfo {author} {\bibfnamefont {S.~G.}\ \bibnamefont
  {Dlamini}}, \bibinfo {author} {\bibfnamefont {J.~T.}\ \bibnamefont
  {Francis}}, \bibinfo {author} {\bibfnamefont {X.}~\bibnamefont {Zhang}},
  \bibinfo {author} {\bibfnamefont {S.~K.}\ \bibnamefont {Özdemir}}, \bibinfo
  {author} {\bibfnamefont {S.~N.}\ \bibnamefont {Chormaic}}, \bibinfo {author}
  {\bibfnamefont {F.}~\bibnamefont {Petruccione}},\ and\ \bibinfo {author}
  {\bibfnamefont {M.~S.}\ \bibnamefont {Tame}},\ }\bibfield  {title} {\bibinfo
  {title} {Probing decoherence in plasmonic waveguides in the quantum regime},\
  }\href@noop {} {\bibfield  {journal} {\bibinfo  {journal} {Physical Review
  Applied}\ }\textbf {\bibinfo {volume} {9}},\ \bibinfo {pages} {024003}
  (\bibinfo {year} {2018})}\BibitemShut {NoStop}%
\bibitem [{\citenamefont {Ebbesen}\ \emph {et~al.}(1998)\citenamefont
  {Ebbesen}, \citenamefont {Lezec}, \citenamefont {Ghaemi}, \citenamefont
  {Thio},\ and\ \citenamefont {Wolff}}]{19eot}%
  \BibitemOpen
  \bibfield  {author} {\bibinfo {author} {\bibfnamefont {T.~W.}\ \bibnamefont
  {Ebbesen}}, \bibinfo {author} {\bibfnamefont {H.~J.}\ \bibnamefont {Lezec}},
  \bibinfo {author} {\bibfnamefont {H.~F.}\ \bibnamefont {Ghaemi}}, \bibinfo
  {author} {\bibfnamefont {T.}~\bibnamefont {Thio}},\ and\ \bibinfo {author}
  {\bibfnamefont {P.~A.}\ \bibnamefont {Wolff}},\ }\bibfield  {title} {\bibinfo
  {title} {Extraordinary optical transmission through sub-wavelength hole
  arrays},\ }\href@noop {} {\bibfield  {journal} {\bibinfo  {journal} {Nature}\
  }\textbf {\bibinfo {volume} {391}},\ \bibinfo {pages} {667} (\bibinfo {year}
  {1998})}\BibitemShut {NoStop}%
\bibitem [{\citenamefont {Rangarajan}\ \emph {et~al.}(2009)\citenamefont
  {Rangarajan}, \citenamefont {Goggin},\ and\ \citenamefont {Kwiat}}]{20spdc}%
  \BibitemOpen
  \bibfield  {author} {\bibinfo {author} {\bibfnamefont {R.}~\bibnamefont
  {Rangarajan}}, \bibinfo {author} {\bibfnamefont {M.}~\bibnamefont {Goggin}},\
  and\ \bibinfo {author} {\bibfnamefont {P.}~\bibnamefont {Kwiat}},\ }\bibfield
   {title} {\bibinfo {title} {Optimizing type-i polarization-entangled
  photons},\ }\href@noop {} {\bibfield  {journal} {\bibinfo  {journal} {Optics
  Express}\ }\textbf {\bibinfo {volume} {17}},\ \bibinfo {pages} {18920}
  (\bibinfo {year} {2009})}\BibitemShut {NoStop}%
\bibitem [{\citenamefont {Bell}(1966)}]{21hiddenvariables}%
  \BibitemOpen
  \bibfield  {author} {\bibinfo {author} {\bibfnamefont {J.}~\bibnamefont
  {Bell}},\ }\bibfield  {title} {\bibinfo {title} {On the problem of hidden
  variables in quantum mechanics},\ }\href@noop {} {\bibfield  {journal}
  {\bibinfo  {journal} {Reviews of Modern Physics}\ }\textbf {\bibinfo {volume}
  {38}},\ \bibinfo {pages} {447} (\bibinfo {year} {1966})}\BibitemShut
  {NoStop}%
\bibitem [{\citenamefont {Clauser}\ \emph {et~al.}(1969)\citenamefont
  {Clauser}, \citenamefont {Horne}, \citenamefont {Shimony},\ and\
  \citenamefont {Holt}}]{22chsh}%
  \BibitemOpen
  \bibfield  {author} {\bibinfo {author} {\bibfnamefont {J.}~\bibnamefont
  {Clauser}}, \bibinfo {author} {\bibfnamefont {M.}~\bibnamefont {Horne}},
  \bibinfo {author} {\bibfnamefont {A.}~\bibnamefont {Shimony}},\ and\ \bibinfo
  {author} {\bibfnamefont {R.}~\bibnamefont {Holt}},\ }\bibfield  {title}
  {\bibinfo {title} {Proposed experiment to test local hidden-variable
  theories},\ }\href@noop {} {\bibfield  {journal} {\bibinfo  {journal}
  {Physical Review Letters}\ }\textbf {\bibinfo {volume} {23}},\ \bibinfo
  {pages} {880} (\bibinfo {year} {1969})}\BibitemShut {NoStop}%
\bibitem [{sup({\natexlab{a}})}]{supplemental1}%
  \BibitemOpen
  \bibfield  {title} {\bibinfo {title} {See supplemental material at [url will
  be inserted by publisher] for discussions on experiments performed in the
  linear regime}}\BibitemShut {NoStop}%
\bibitem [{sup({\natexlab{b}})}]{supplemental2}%
  \BibitemOpen
  \bibfield  {title} {\bibinfo {title} {See supplemental material at [url will
  be inserted by publisher] for discussions on the operation of the hole array
  structure in the highly dispersive regime}}\BibitemShut {NoStop}%
\bibitem [{\citenamefont {Puech}\ \emph {et~al.}(1995)\citenamefont {Puech},
  \citenamefont {Henari}, \citenamefont {Blau}, \citenamefont {Duff},\ and\
  \citenamefont {Schmid}}]{23dephasing}%
  \BibitemOpen
  \bibfield  {author} {\bibinfo {author} {\bibfnamefont {K.}~\bibnamefont
  {Puech}}, \bibinfo {author} {\bibfnamefont {F.~Z.}\ \bibnamefont {Henari}},
  \bibinfo {author} {\bibfnamefont {W.~J.}\ \bibnamefont {Blau}}, \bibinfo
  {author} {\bibfnamefont {D.}~\bibnamefont {Duff}},\ and\ \bibinfo {author}
  {\bibfnamefont {G.}~\bibnamefont {Schmid}},\ }\bibfield  {title} {\bibinfo
  {title} {Investigation of the ultrafast dephasing time of gold nanoparticles
  using incoherent light},\ }\href@noop {} {\bibfield  {journal} {\bibinfo
  {journal} {Chemical Physics Letters}\ }\textbf {\bibinfo {volume} {247}},\
  \bibinfo {pages} {13} (\bibinfo {year} {1995})}\BibitemShut {NoStop}%
\end{thebibliography}
\providecommand{\noopsort}[1]{}\providecommand{\singleletter}[1]{#1}%

\end{document}